\documentclass[10pt,preprint]{aastex}

\slugcomment{Not to appear in Nonlearned J., 45.}

\shorttitle{ALMA observations of N83C in SMC}
\shortauthors{Muraoka et al.}

\begin{document}


\title{ALMA observations of N83C in the early stage of star formation in the Small Magellanic Cloud}


\author{Kazuyuki Muraoka\altaffilmark{1}, Aya Homma\altaffilmark{1}, Toshikazu Onishi\altaffilmark{1}, Kazuki Tokuda\altaffilmark{1,2}, Ryohei Harada\altaffilmark{1}, Yuuki Morioka\altaffilmark{1}, Sarolta Zahorecz\altaffilmark{1,2}, Kazuya Saigo\altaffilmark{2}, Akiko Kawamura\altaffilmark{2}, Norikazu Mizuno\altaffilmark{2,3}, Tetsuhiro Minamidani\altaffilmark{3,4}, Erik Muller\altaffilmark{2}, Yasuo Fukui\altaffilmark{5}, Margaret Meixner\altaffilmark{6,7}, Remy Indebetouw\altaffilmark{8,9}, Marta Sewi{\l}o\altaffilmark{10}}

\and
\author{Alberto Bolatto\altaffilmark{11}}

\altaffiltext{1}{Department of Physical Science, Graduate School of Science, Osaka Prefecture University, 1-1 Gakuen-cho, Naka-ku, Sakai, Osaka 599-8531, Japan}
\altaffiltext{2}{Chile Observatory, National Astronomical Observatory of Japan, National Institutes of Natural Science, 2-21-1 Osawa, Mitaka, Tokyo 181-8588, Japan}
\altaffiltext{3}{Department of Astronomical Science, School of Physical Science, SOKENDAI (The Graduate University of Advanced Studies), 2-21-1 Osawa, Mitaka, Tokyo 181-8588, Japan}
\altaffiltext{4}{Nobeyama Radio Observatory, National Astronomical Observatory of Japan, National Institutes of Natural Science, 462-2 Nobeyama, Minamimaki-mura, Minamisaku-gun, Nagano 384-1305, Japan}
\altaffiltext{5}{Department of Astrophysics, Nagoya University, Chikusa-ku, Nagoya 464-8602, Japan}
\altaffiltext{6}{The Johns Hopkins University, Department of Physics and Astronomy, 366 Bloomberg Center, 3400 N. Charles Street, Baltimore, MD 21218, USA}
\altaffiltext{7}{Space Telescope Science Institute, 3700 San Martin Drive, Baltimore, MD 21218, USA}
\altaffiltext{8}{Department of Astronomy, University of Virginia, P.O. Box 400325, Charlottesville, VA 22904, USA}
\altaffiltext{9}{National Radio Astronomy Observatory, 520 Edgemont Road, Charlottesville, VA 22903, USA}
\altaffiltext{10}{NASA Goddard Space Flight Center, 8800 Greenbelt Road, Greenbelt, MD 20771, USA}
\altaffiltext{11}{Department of Astronomy, University of Maryland, College Park, MD 20742, USA}



\begin{abstract}
We have performed Atacama Large Millimeter/submillimeter Array (ALMA) observations in $^{12}$CO($J=2-1$), $^{13}$CO($J=2-1$), C$^{18}$O($J=2-1$),
$^{12}$CO($J=3-2$), $^{13}$CO($J=3-2$), and CS($J=7-6$) lines toward the active star-forming region N83C in the Small Magellanic Cloud (SMC), whose metallicity is about one-fifth of the Milky Way (MW).
The ALMA observations first reveal subparsec-scale molecular structures in $^{12}$CO($J=2-1$) and $^{13}$CO($J=2-1$) emissions.
We found strong CO peaks associated with young stellar objects (YSOs) identified by the $Spitzer$ $Space$ $Telescope$,
and we also found that overall molecular gas is distributed along the edge of the neighboring H$\,${\sc ii} region.
We derived a gas density of $\sim 10^4$ cm$^{-3}$ in molecular clouds associated with YSOs based on the virial mass estimated from the $^{12}$CO($J=2-1$) emission.
This high gas density is presumably due to the effect of the H$\,${\sc ii} region under the low-metallicity (and accordingly small-dust content) environment in the SMC;
far-UV radiation from the H$\,${\sc ii} region can easily penetrate and photodissociate the outer layer of $^{12}$CO molecules in the molecular clouds,
and thus only the innermost parts of the molecular clouds are observed even in $^{12}$CO emission.
We obtained the CO-to-H$_2$ conversion factor $X_{\rm CO}$ of $7.5 \times 10^{20}$ cm$^{-2}$ (K km s$^{-1}$)$^{-1}$ in N83C based on virial masses and CO luminosities,
and it is four times larger than that in the MW, 2 $\times 10^{20}$ cm$^{-2}$ (K km s$^{-1}$)$^{-1}$.
We also discuss the difference in the nature between two high-mass YSOs, each of which is associated with a molecular clump with a mass of about a few $\times 10^3 M_{\odot}$.

\end{abstract}

\keywords{ISM: clouds --- ISM: molecules --- stars: formation}

\section{Introduction}

Metallicity in the interstellar medium is one of the key parameters
to control star-formation processes and molecular gas properties.
In particular, an environment with low metallicity (and accordingly small-dust content) is important
because it causes a reduction of the shielding against far-UV radiation from massive stars,
which affects the formation, structure, and physical properties of giant molecular clouds (GMCs) as well as the star-formation process.

The Small Magellanic Cloud (SMC), which is classified as a dwarf irregular galaxy, is an ideal target to investigate molecular gas structures and star-formation processes
under an environment with low metallicity and small-dust content because its metallicity is about one-fifth of the Sun (\citealt{dufour1984}; \citealt{kurt1999}; \citealt{pagel2003})
and the dust-to-gas ratio is 17 times smaller than that in the Milky Way (MW; \cite{koornneef1984}.
In addition, its proximity ($\sim$ 60 kpc; \citealt{hilditch2005}) enables us to perform high-spatial-resolution observations even at millimeter to submillimeter wavelengths.

Molecular clouds in the SMC have been observed in the $^{12}$CO($J=1-0$), $^{13}$CO($J=1-0$), $^{12}$CO($J=2-1$), and $^{13}$CO($J=2-1$) lines
using the Swedish-ESO Submillimeter Telescope at angular resolutions of 45$''$ ($\sim$ 13 pc) for CO($J=1-0$) and 23$''$ ($\sim$ 7 pc) for CO($J=2-1$),
and intensity ratios among these molecular lines have been examined by many authors (e.g., \citealt{israel1993}; \citealt{rubio1993}; \citealt{lequeux1994}; \citealt{rubio1996}; \citealt{israel2003}).
In addition, \cite{mizuno2001} observed the SMC in the $^{12}$CO($J=1-0$) line at an angular resolution of $2$\rlap.{'}$6$ using the NANTEN telescope,
and they identified 21 GMCs whose masses are $\sim 10^4$ to $10^6 M_{\odot}$.
They found a good spatial correlation between the GMCs and H$\,${\sc ii} regions or young clusters, indicating that cluster formation is ongoing in these GMCs.

Here, we focus on the N83 region, which is located in the southeast wing of the SMC and is an isolated, yet relatively active star-forming region.
The H$\,${\sc ii} region contains the stellar association NGC456 and a possible supernova remnant \citep{haberl2000}.
Molecular gas delineates the edge of the H$\,${\sc ii} region \citep{bolatto2003},
and this indicates that the molecular cloud may be formed or compressed by the expanding shell.
In addition, warm ($\sim$ 40 K) molecular gas exists in N83 that might be heated by the neighboring H$\,${\sc ii} region (\citealt{bolatto2003}; \citealt{bolatto2005}).
The CO-to-H$_{2}$ conversion factor ($X_{\rm CO}$) in N83 was estimated to be ($7.0 \pm 3.4$) $\times 10^{20}$ cm$^{-2}$ (K km s$^{-1}$)$^{-1}$ \citep{bolatto2003}.

In particular, \cite{bolatto2003} found that the south clump N83C shows the most intense CO($J=2-1$) emission in the N83 region,
and its $X_{\rm CO}$ is lower than that in other regions of N83.
They argued that $X_{\rm CO}$ decreases and reaches nearly Galactic values in CO-bright regions such as N83C (see also \citealt{glover2012}).
However, these observations could not fully resolve molecular clouds (e.g., \citealt{israel2003}),
and thus smaller structures such as cores and filaments may exist.
For such unresolved clouds, one cannot reveal the detailed evolutionary processes of molecular clouds associated with young stellar objects (YSOs).
In addition, the estimated $X_{\rm CO}$ might be uncertain, due to the uncertainties of the cloud geometry and the resultant virial mass.
In order to resolve smaller molecular structures and to obtain more accurate $X_{\rm CO}$ values in the SMC,
further observations with higher angular resolution are indispensable.

In this paper, we present the results of Atacama Large Millimeter/submillimeter Array (ALMA) observations toward N83C.
Firstly, we reveal the distribution of molecular gas associated with YSOs based on the high-angular-resolution maps with ALMA.
Then, we derive various properties of molecular clouds: virial mass, average gas density, kinetic temperature, and $X_{\rm CO}$.
Finally, we comprehensively discuss molecular gas properties and their relation to star formation in N83C.

\section{Observations}

ALMA Cycle 2 observations toward N83C were carried out in Band~6 (211 -- 275~GHz) and Band~7 (275 -- 373~GHz)
with the main array 12~m antennas, the Atacama Compact Array (ACA) 7~m antennas, and total power (TP) 12~m antennas
between 2014 May and 2015 September.

The observations were centered on ($\alpha_{\rm J2000}$, $\delta_{\rm J2000}$) = ($01^h 14^m 05\rlap.{^s}414$, $-73^{\circ} 17' 03\rlap.{''}908$).
The target molecular lines are $^{12}$CO($J=2-1$), $^{13}$CO($J=2-1$), and C$^{18}$O($J=2-1$) in Band~6,
and $^{12}$CO($J=3-2$), $^{13}$CO($J=3-2$), and CS($J=7-6$) in Band~7
with a bandwidth of 117.19 MHz (30.5 kHz $\times$ 3840 channels).
Note that we used a spectral window for the observations of the continuum emission with a bandwidth of 1875.0 MHz (488.3 kHz $\times$ 3840 channels) in Band~6.
The maximum baseline of the 12~m array is 348.5~m and that of ACA is 48.9~m.

The data were reduced using the Common Astronomy Software Application (CASA) package.
We applied natural weighting for both Band~6 and Band~7 data,
which provided synthesized beam sizes of 1$\rlap.{''}$72 $\times$ 1$\rlap.{''}$37 (0.50~pc $\times$ 0.40~pc at 60~kpc)
and 4$\rlap.{''}$64 $\times$ 4$\rlap.{''}$33 (1.35 pc $\times$ 1.23 pc), respectively.
Note that we only used ACA and TP data in Band~7 because we could not obtain the main-array data,
and thus its beam size is larger than that in Band~6.
The rms noise levels of molecular lines at the velocity resolution of 0.2 km s$^{-1}$
are 17 mJy beam$^{-1}$ in Band~6 and 148 mJy beam$^{-1}$ in Band~7, respectively.

\section{Results and Analyses}

\subsection{Maps}

Figures~1(a) and (b) show integrated intensity maps in $^{12}$CO($J=2-1$) and $^{13}$CO($J=2-1$) emission of N83C.
The global distribution of $^{12}$CO($J=2-1$) emission is quite similar to that of $^{13}$CO($J=2-1$) emission.
Strong peaks of CO emission correspond to the position of bright YSOs\footnote{We assigned the source ID to each YSO in alphabetical order
according to the $^{12}$CO($J=2-1$) peak temperature (see Table 1).} (A and B), identified based on the $Spitzer$ $Space$ $Telescope$ data (see also Figure 1(c)).
Thus these YSOs are born within such CO-bright molecular clouds.
We can see faint local peaks at sources C and D, but no CO emission is found at the source E.
Figure~1(d) shows that local peaks of H$\alpha$ emission are associated with all of the sources, A to E.
This implies that all of the YSOs are beginning to ionize the surrounding molecular gas.
Overall molecular gas is distributed along the edge of the H$\,${\sc ii} region.

In Figures 2(a) and (b), we found a similar trend in the $J=3-2$ transition as in the $J=2-1$ transition (Figures 1(a) and (b));
that is, the global distributions of $^{12}$CO($J=3-2$) and $^{13}$CO($J=3-2$) are quite similar to each other,
and strong peaks are observed at sources A and B in both CO emissions.
Figures~2(c) and (d) show integrated intensity maps in the CS($J=7-6$) and C$^{18}$O($J=2-1$)\footnote{We used only ACA and TP data
but excluded main-array data for C$^{18}$O($J=2-1$) emissions because the main-array data degrade the resultant signal-to-noise ratio of the C$^{18}$O($J=2-1$) map.
Therefore, its beam size (8$\rlap.{''}$25 $\times$ 6$\rlap.{''}$06) is larger than that of the $^{12}$CO($J=2-1$) map (1$\rlap.{''}$72 $\times$ 1$\rlap.{''}$37).} emission of N83C.
Although both lines can trace dense cores in molecular clouds, we find different features in their respective maps.
The CS($J=7-6$) intensity is stronger in source A than in source B,
whereas the C$^{18}$O($J=2-1$) peak is only observed at source B (see Section 4.2 in detail).
The peak temperatures of each emission in sources A to E are summarized in Table 1.

We estimate the virial mass ($M_{\rm vir}$) of molecular clouds associated with sources A and B as follows \citep{solomon1987}:
\begin{eqnarray}
M_{\rm vir} = 1040 \sigma_v^2 R \,\, ,
\end{eqnarray}
where $R$ is the radius of a molecular cloud in pc, and $\sigma_v$ is the cloud velocity dispersion in km s$^{-1}$.
Here, $R$ was determined as the effective radius of the area whose CO intensity exceeds half the peak,
and $\sigma_v$ was calculated from the gaussian fitting of the CO line integrated over the area.
Both $R$ and $\sigma_v$ are determined from the $^{12}$CO($J=2-1$) map and the $^{13}$CO($J=2-1$) map individually, as shown in Table 2.
The resultant $M_{\rm vir}$ values are $\sim 10^3 M_{\odot}$.

Using the $M_{\rm vir}$, we derive the average gas density $n({\rm H}_2)$ in the molecular clumps as follows:
\begin{eqnarray}
n({\rm H}_2) = \frac{M_{\rm vir} / (\mu_{\rm H_2} \times m_{\rm H})}{4 \pi R^3 / 3} \,\, ,
\end{eqnarray}
where $\mu_{\rm H_2} \sim 2.7$ is the mean molecular weight per H$_2$ molecule (including the contribution by He), and $m_{\rm H}$ is the atomic hydrogen mass.
The estimated $n({\rm H}_2)$ values are $\sim 10^4$ cm$^{-3}$ (see Table 2).

\subsection{Deriving gas density and kinetic temperature of molecular clouds}

Here, we estimate gas density and kinetic temperature of molecular clouds using line ratios obtained from ALMA observations
based on the Large Velocity Gradient (LVG) approximation (e.g., \citealt{goldreich1974}; \citealt{scoville1974}).

Firstly, we examined spatial distributions of $^{13}$CO($J=3-2$)/$^{12}$CO($J=3-2$) and
$^{13}$CO($J=3-2$)/$^{13}$CO($J=2-1$) ratios as shown in Figures 3(a) and (b).
We found local peaks corresponding to sources A and B in the $^{13}$CO($J=3-2$)/$^{12}$CO($J=3-2$) ratio map,
whereas such local peaks are not observed in the $^{13}$CO($J=3-2$)/$^{13}$CO($J=2-1$) ratio map.

Then, we carried out the LVG calculation assuming the following input parameters:
the molecular abundance $Z$($^{12}$CO) = [$^{12}$CO]/[H$_2$] of $8.0 \times 10^{-6}$,
the [$^{12}$CO]/[$^{13}$CO] abundance ratio of 50, and the velocity gradient $dv/dr$ of 2.0 km s$^{-1}$ pc$^{-1}$.
The assumed $Z$($^{12}$CO) is based on that in the Large Magellanic Cloud (LMC);
\citet{mizuno2010} assumed $Z$($^{12}$CO) of $1.6 \times 10^{-5}$ in the LMC,
whose metallicity is twice that in the SMC.
Thus we assumed the twice smaller $Z$($^{12}$CO) of $8.0 \times 10^{-6}$.
The assumed [$^{12}$CO]/[$^{13}$CO] of 50 is the same as that in the LMC \citep{mizuno2010}.
The assumed $dv/dr$ of 2.0 km s$^{-1}$ pc$^{-1}$ is based on the size and the velocity dispersion of the molecular clump in source A (see Table 2).

Figures~3(c) and (d) show spatial distributions of molecular gas density and kinetic temperature, respectively.
We found gas densities of $\sim 1.5 \times 10^4$ cm$^{-3}$ and kinetic temperatures of 30 -- 50 K at sources A and B.
The derived gas density based on the LVG calculation is consistent with that calculated from the virial mass.
This implies that each molecular clump is gravitationally bounded.
Note that we found a linear feature of high $T_{\rm K}$ ($\sim$ 90 K) from southeast to northwest in Figure 3(d).
This is due to the breakdown of the one-zone model in the LVG approximation by the overlap of molecular gas with multiple velocity components.

\section{Discussion}

\subsection{$X_{\rm CO}$ in the SMC}

So far, $X_{\rm CO}$ in molecular clouds has been estimated mainly using virial masses and CO luminosities.
It is reported that the standard $X_{\rm CO}$ value in the MW disk is 2 $\times 10^{20}$ cm$^{-2}$ (K km s$^{-1}$)$^{-1}$ (\citealt{bolatto2013} and references therein).
However, $X_{\rm CO}$ may vary depending on the mass and the definition of molecular clouds;
for example, smaller clouds tend to have higher $X_{\rm CO}$, i.e., $X_{\rm CO} \propto L_{\rm CO}^{-0.2}$ \citep{bolatto2013}.
Our ALMA observations spatially resolved molecular clumps associated with YSOs (sources A and B);
each clump is an active site of star formation and is likely to have nearly a round shape.
This indicates that each clump is likely to be gravitationally bounded, which is also estimated from the LVG analysis in Section 3.2, with nearly a spherical shape.
Thus we derive the ``spatially-resolved'' $X_{\rm CO}$ of sources A and B in N83C and compare it with that obtained in earlier studies for SMC and the MW.
The $X_{\rm CO}$ value can be calculated from the following equation:
\begin{eqnarray}
X_{\rm CO} = \frac{M_{\rm vir} / (\mu_{\rm H_2} \times m_{\rm H})}{L_{\rm CO}} \,\, ,
\end{eqnarray}
where $L_{\rm CO}$ is the $^{12}$CO($J=1-0$) luminosity of the molecular cloud.
In order to calculate $L_{\rm CO}$ from our $^{12}$CO($J=2-1$) data, we assume the $^{12}$CO($J=2-1$)/$^{12}$CO($J=1-0$) ratio of 0.9 according to \cite{bolatto2003}.
This ratio of 0.9 is consistent with the massive star-forming region in the MW (e.g., \citealt{nishimura2015}).
For source A, $M_{\rm vir}$ of $2.5 \times 10^3 M_{\odot}$ and $L_{\rm CO}$ of 186 K km s$^{-1}$ pc$^2$ yield
$X_{\rm CO}$ of $6.2 \times 10^{20}$ cm$^{-2}$ (K km s$^{-1}$)$^{-1}$.
In addition, we obtained a similar $X_{\rm CO}$ from source B, $8.7 \times 10^{20}$ cm$^{-2}$ (K km s$^{-1}$)$^{-1}$,
from $M_{\rm vir}$ of $1.5 \times 10^3 M_{\odot}$ and $L_{\rm CO}$ of 81 K km s$^{-1}$ pc$^2$.
The averaged $X_{\rm CO}$ of $7.5 \times 10^{20}$ cm$^{-2}$ (K km s$^{-1}$)$^{-1}$ is four times higher than that in the MW disk, 
which is in good agreement with that in N83 obtained by \cite{bolatto2003}, ($7.0 \pm 3.4$) $\times 10^{20}$ cm$^{-2}$ (K km s$^{-1}$)$^{-1}$.

\subsection{Properties of molecular gas and star formation in N83C}

We summarize the properties of the molecular gas based on our ALMA observations
and discuss their relation to star formation in N83C.

The derived gas density of $\sim 10^4$ cm$^{-3}$ is significantly higher than the typical gas density of $2 \times 10^3$ cm$^{-3}$ in the Orion A/B molecular clouds \citep{nishimura2015},
even though both gas densities are derived from the same molecular line, $^{12}$CO($J=2-1$).
Such a difference in derived gas densities is presumably due to the low metallicity in the SMC.
The low metallicity corresponds to the small-dust content in molecular clouds, which causes the reduction of the shielding against far-UV radiation.
In this condition, external far-UV radiation from the H$\,${\sc ii} region associated with N83C easily penetrates and photodissociates the outer layer of the $^{12}$CO molecule in the molecular clouds.
Thus we can only observe the innermost parts of the molecular clouds, whose gas density is $\sim 10^4$ cm$^{-3}$, even in $^{12}$CO emission.
It is to be noted that theoretical simulations examined by \citet{glover2012} support our results:
their simulations suggest that diffuse CO emission becomes unobservable, but dense gas continues to produce observable CO emission as metallicity decreases.
The observational fact that even $^{12}$CO emission traces the high-density ($\sim 10^4$ cm$^{-3}$) regions in N83C just corresponds to this situation.

Then, we consider the meaning of the derived kinetic temperature of molecular gas, 30 -- 50 K.
\cite{bolatto2003} suggested that warm ($\sim$ 40 K) clumps exist in some regions of N83 whose $^{12}$CO($J=2-1$)/$^{12}$CO($J=1-0$) ratios exceed 2.
Although the $^{12}$CO($J=2-1$)/$^{12}$CO($J=1-0$) ratio of 0.9 in N83C is not so high (\citealt{bolatto2003}; \citealt{bolatto2005}),
it is possible that the global gas temperature exceeds 30 K in N83C considering that the bright H$\,${\sc ii} region, which is a likely heating source, spreads over N83.
In addition, the kinetic temperature of molecular gas is consistent with the dust temperature of 35~K \citep{takekoshi2017}.
Recent observational studies in the MW can also support our results:
\cite{nishimura2015} reported that the kinetic temperature in Orion A/B molecular clouds is 20 -- 50 K, which is similar to that in the N83C,
although the typical temperature of molecular clouds in the MW is 10 -- 20 K.
Considering that both N83C and Orion A/B molecular clouds are associated with bright H$\,${\sc ii} regions,
it is suggested that their temperature increase is due to the heating by far-UV radiation.
However, \cite{nishimura2015} also found a temperature gradient in Orion A/B molecular clouds;
the kinetic temperature decreases with distance from the H$\,${\sc ii} region,
while such a temperature gradient extending over more than a few parsecs is not observed in N83C with the spatial resolution of $\sim$ 1 pc.
Thus there is a possibility that the molecular gas temperature in N83C is intrinsically high before the generation of the neighboring H$\,${\sc ii} region.

Finally, we discuss the difference in the nature between sources A and B.
As described in Section 3.1, the CS($J=7-6$) intensity is stronger in source A than in source B, whereas the C$^{18}$O($J=2-1$) peak is only observed at source B.
This can be explained by the effects of far-UV radiation from the internal YSOs.
Strong far-UV radiation can photochemically enhance CS emission (e.g., \citealt{zhou1991}; \citealt{jansen1994}; \citealt{sternberg1995})
but dissociate the C$^{18}$O molecules (e.g., \citealt{shimajiri2014}).
Considering the luminosity of source A, 20140 $L_{\odot}$, is three times larger than that of source B, 6899 $L_{\odot}$ \citep{kamath2014},
our observational results suggest that strong far-UV radiation from bright YSOs enhances CS emission but dissociates C$^{18}$O molecule in source A,
while such effects do not work in source B because of its lower brightness.
However, we found an opposite trend in the distribution of H$\alpha$ flux.
As shown in the left panel of Figure 4, the peak of H$\alpha$ emission is not observed at source A, where H$\alpha$ emission is weak,
although the strong H$\alpha$ peak is observed at source B.
In order to reveal the cause of the weak H$\alpha$ emission in source A,
we examined the spatial comparison between H$\alpha$ flux and the Band~6 continuum emission, which corresponds to emission from cold dust, as shown in the right panel of Figure 4.
The strongest peak of the Band~6 continuum emission is coincident with the position of source A, and a possible peak is found at source B.
This implies a very high density (and presumably unresolved) gas clump at source A and there seems to be a deeply embedded and very cold YSO therein,
although the molecular gas density estimated from the LVG calculation is similar between sources A and B ($1.5 \times 10^4$ cm$^{-3}$).
Note that the source A is seen at the $BVI$ bands \citep{kamath2014} in spite of the existence of the high-density clump,
indicating that source A is likely a composite with a YSO deeply embedded.
We also note the H$\alpha$ peak located to the south of source A, which is labeled as source F in Figure 4.
It seems that the $Spitzer$ map cannot resolve sources A and F spatially, as shown in Figure 1(c).
In addition, the CO clump at source A elongates toward source F.
These observational facts indicate that source F is also a YSO as are sources A to E.

\section{Summary and Conclusions}

We have performed ALMA Cycle 2 observations in the $^{12}$CO($J=2-1$), $^{12}$CO($J=2-1$), C$^{18}$O($J=2-1$),
$^{12}$CO($J=3-2$), $^{13}$CO($J=3-2$) and CS($J=7-6$) lines toward N83C.
We summarize this work is as follows:

\begin{enumerate}
\item We have successfully obtained subparsec-scale molecular structures in $^{12}$CO($J=2-1$) and $^{13}$CO($J=2-1$) emissions, and we found strong CO peaks associated with YSOs.
Overall molecular gas is distributed along the edge of the H$\,${\sc ii} region.

\item We derived a gas density of $\sim 10^4$ cm$^{-3}$ of molecular clouds associated with YSOs based on the virial mass estimated from $^{12}$CO($J=2-1$) emission as well as the LVG calculation.
Since the outer layer of $^{12}$CO molecules in the molecular clouds might be photodissociated, we can only observe the innermost parts of the molecular clouds with high gas density.

\item We obtained $X_{\rm CO}$ of $7.5 \times 10^{20}$ cm$^{-2}$ (K km s$^{-1}$)$^{-1}$ for star-forming clumps in N83C based on virial masses and CO luminosities,
a number that is four times larger than that in the MW, 2 $\times 10^{20}$ cm$^{-2}$ (K km s$^{-1}$)$^{-1}$.
Note that clouds with $L_{\rm CO}$ much smaller than $10^5$ K km s$^{-1}$ pc$^2$ have $X_{\rm CO}$ a few times larger than this standard value, even in the MW \citep{bolatto2013}.

\item We derived the kinetic temperature of the molecular clouds of 30 -- 50 K.
There are two possibilities to explain such a high temperature: one is the heating from the neighboring H$\,${\sc ii} region,
and the other is that the molecular gas temperature in N83C is intrinsically high before the generation of the neighboring H$\,${\sc ii} region.

\item We discussed the difference in the nature between two high-mass YSOs, each of which is associated with a molecular clump with a mass of about a few $\times 10^3 M_{\odot}$.
The luminosities of YSOs are brighter in source A than in source B, whereas H$\alpha$ emission is brighter in source B than in source A.
Considering the strong peak of the Band~6 (230 GHz) continuum emission at source A, a very high-density gas clump may exist.

\end{enumerate}

\begin{table}
\begin{center}
Table~1.\hspace{4pt}Properties in Each Molecular Line
\begin{tabular}{cccccccccc}
\hline \hline \\[-4mm]
Molecular line & \multicolumn{5}{c}{Peak Temperature in Each Source} & Beam Size & Position Angle & \multicolumn{2}{c}{Rms Noise Level} \\
  & \multicolumn{5}{c}{(K)} & (arcsec) & (degree) & (mJy beam$^{-1})$ & (mK) \\
  & A & B & C & D & E & & & & \\
\hline \\[-4mm]
$^{12}$CO($J$=2-1) & 24.4  & 21.7 & 18.7  & 7.1      & 5.6      & 1\farcs72 $\times$ 1\farcs37 & 44$^\circ$  & 17  & 162  \\ 
$^{13}$CO($J$=2-1) & 7.2   & 7.9  & 4.3   & 1.9      & 1.6      & 1\farcs75 $\times$ 1\farcs40 & 46$^\circ$  & 19  & 188  \\
$^{12}$CO($J$=3-2) & 18.6  & 14.6 & 13.8  & 3.5      & 1.7      & 4\farcs64 $\times$ 4\farcs33 & -33$^\circ$ & 148 & 68   \\
$^{13}$CO($J$=3-2) & 4.1   & 4.2  & 2.8   & 0.52     & 0.37     & 4\farcs93 $\times$ 4\farcs62 & 61$^\circ$  & 174 & 79   \\
CS($J$=7-6)        & 0.58  & 0.49 & 0.36  & $\cdots$ & $\cdots$ & 4\farcs75 $\times$ 4\farcs06 & 56$^\circ$  & 113 & 64   \\
C$^{18}$O($J$=2-1) & 0.077 & 0.10 & 0.079 & $\cdots$ & $\cdots$ & 8\farcs25 $\times$ 6\farcs06 & 86$^\circ$  & 37  & 16   \\
\hline \\[-3mm]
\end{tabular}\\
{\footnotesize References for YSOs: A and B are identified by \cite{kamath2014}, and C, D, and E are by \cite{sewilo2013}.
}
\end{center}
\end{table}

\begin{table}
\begin{center}
Table~2.\hspace{4pt}Virial Masses, H$_2$ Densities, and CO Luminosities of CO Clumps in N83C
\begin{tabular}{ccccccc}
\hline \hline \\[-4mm]
Source & molecular line & $\sigma_v$    & $R$  & $M_{\rm vir}$ & H$_2$ density & $L_{\rm CO}$ \\
       &                & (km s$^{-1}$) & (pc) & ($M_{\odot}$) & (cm$^{-3}$)   & (K km s$^{-1}$ pc$^2$) \\
\hline \\[-4mm]
A & $^{12}$CO($J$=2-1) & 1.69 & 0.8 & 2.5 $\times$ 10$^3$ & 1.5 $\times$ 10$^4$ & 186 \\
  & $^{13}$CO($J$=2-1) & 1.33 & 0.7 & 1.4 $\times$ 10$^3$ & 1.2 $\times$ 10$^4$ & $\cdots$ \\
\hline
B & $^{12}$CO($J$=2-1) & 1.60 & 0.6 & 1.5 $\times$ 10$^3$ & 2.8 $\times$ 10$^4$ & 81 \\
  & $^{13}$CO($J$=2-1) & 0.94 & 0.6 & 5.8 $\times$ 10$^2$ & 7.9 $\times$ 10$^3$ & $\cdots$ \\
\hline \\[-3mm]
\end{tabular}\\
{\footnotesize $L_{\rm CO}$ means the CO($J=1-0$) luminosity assuming the CO($J=2-1$)/CO($J=1-0$) intensity ratio of 0.9.
}
\end{center}
\end{table}


\begin{figure*}[ht!]
\begin{center}
\plotone{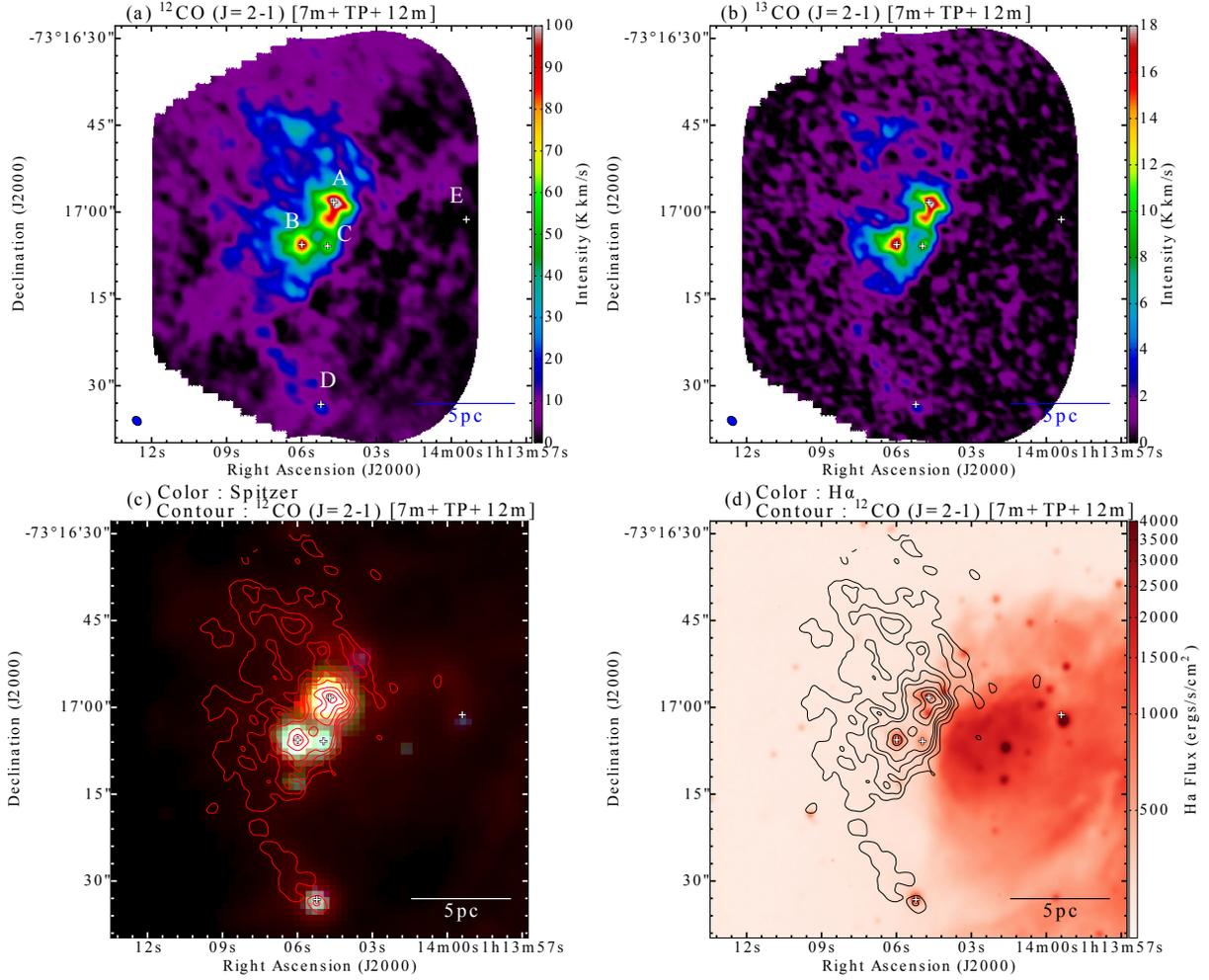}
\caption{(a) Integrated intensity map in $^{12}$CO($J=2-1$) emission of N83C. Crosses show the positions of bright sources (YSOs)
identified by the $Spitzer$ $Space$ $Telescope$ (\citealt{sewilo2013}; \citealt{kamath2014}).
The synthesized beam is shown in the lower left corner.
(b) Integrated intensity map in $^{13}$CO($J=2-1$) emission.
(c) Three-color composite $Spitzer$/IRAC image (blue: 3.6 $\mu$m, green: 4.5 $\mu$m, and red: 8 $\mu$m)
superposed on the integrated intensity map in $^{12}$CO($J=2-1$) emission (contour).
The contour levels are 10, 20, 30, 40, 60, 80, and 100 K km s$^{-1}$.
(d) H$\alpha$ flux map (color, \citealt{li2017}) superposed on the integrated intensity map in $^{12}$CO($J=2-1$) emission (contour).
The contour levels are the same as in (c).
}
\end{center}
\end{figure*}

\begin{figure*}[ht!]
\plotone{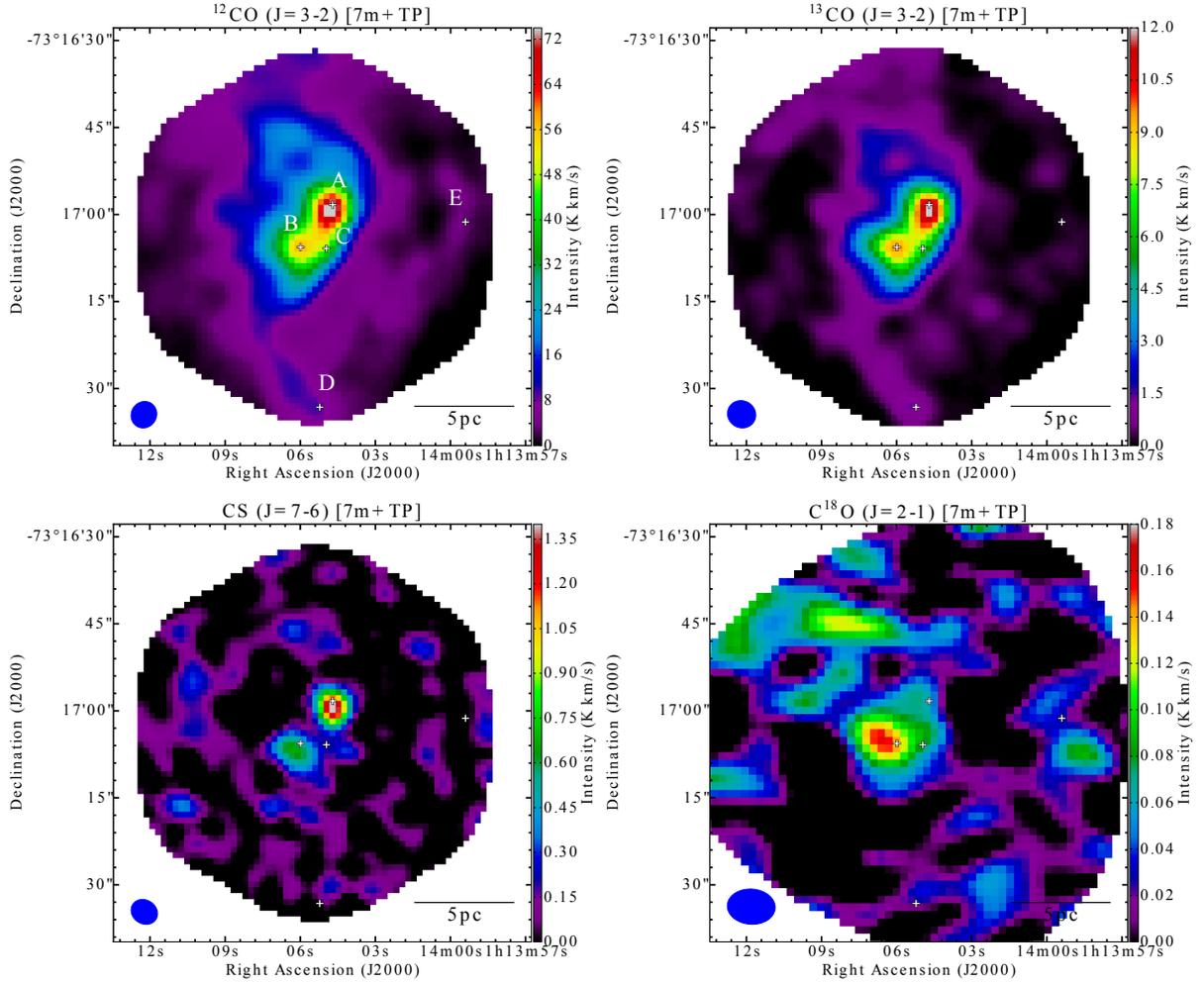}
\caption{Integrated intensity maps in $^{12}$CO($J=3-2$), $^{13}$CO($J=3-2$), CS($J=7-6$), and C$^{18}$O($J=2-1$) emission of N83C.
The synthesized beam is shown in the lower left corner of each map.
}
\end{figure*}

\begin{figure*}[ht!]
\plotone{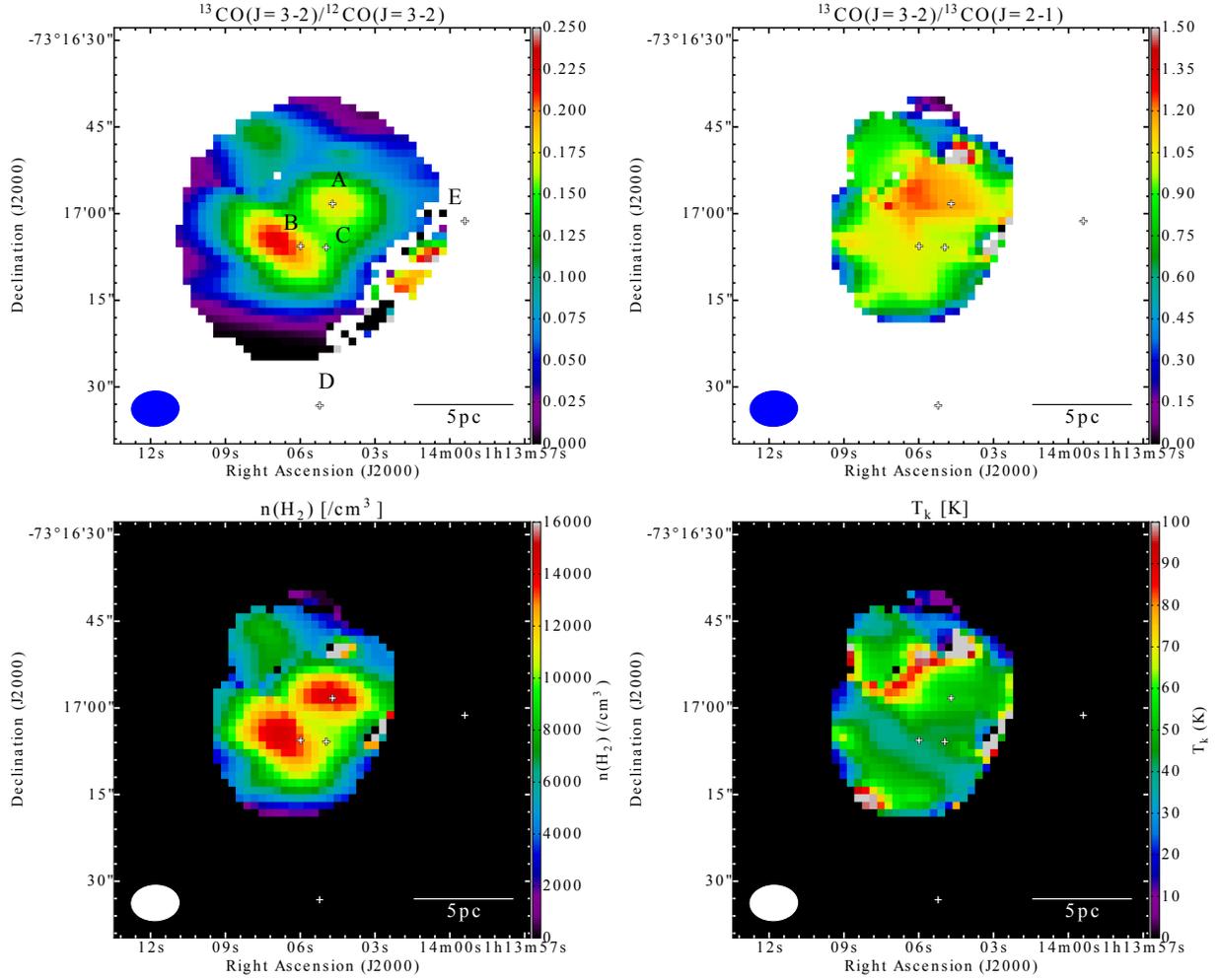}
\caption{Top: spatial distributions of the $^{13}$CO($J=3-2$)/$^{12}$CO($J=3-2$) and the $^{13}$CO($J=3-2$)/$^{13}$CO($J=2-1$) intensity ratios of N83C.
Bottom: spatial distributions of molecular gas density $n({\rm H}_2)$ and kinetic temperature $T_{\rm K}$ of N83C derived from the LVG calculation.
}
\end{figure*}

\begin{figure*}[ht!]
\plotone{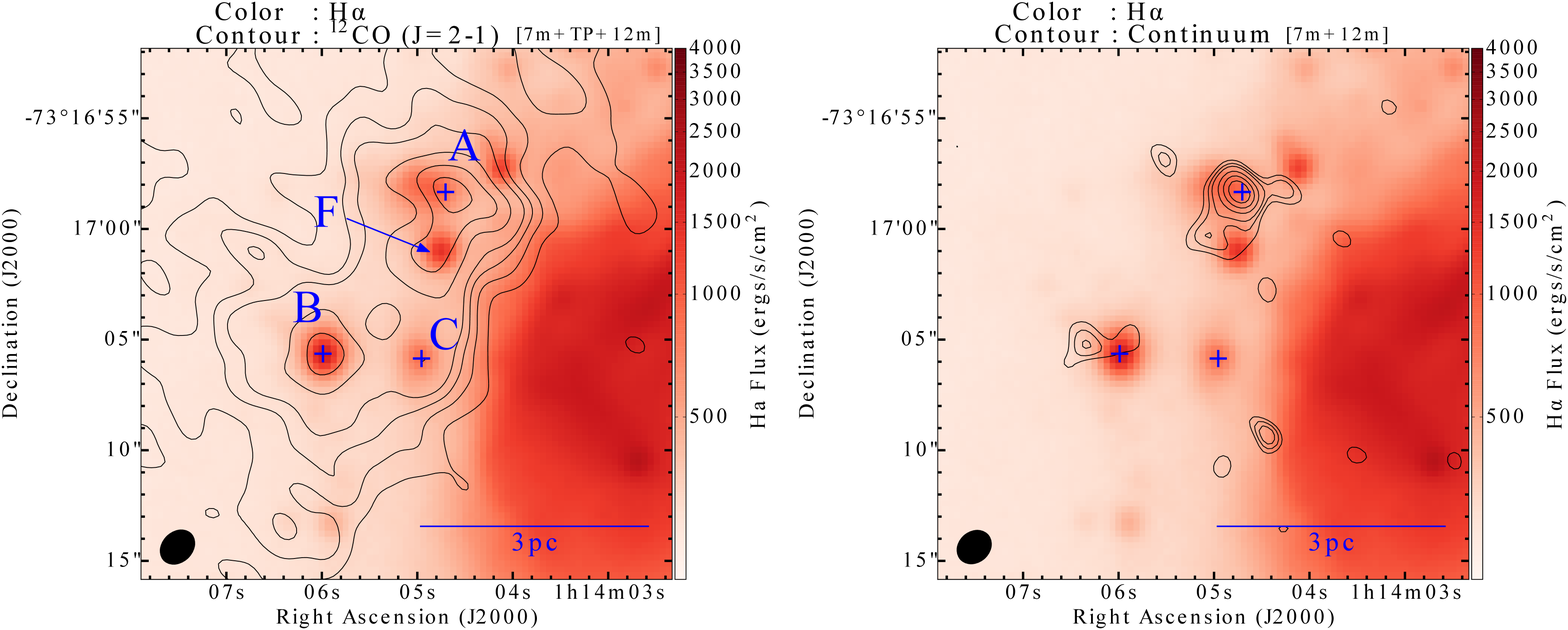}
\caption{Left: enlarged view of Figure 1(d): integrated intensity map in $^{12}$CO($J=2-1$) emission (contour) superposed on H$\alpha$ flux map (color)
at sources A to C and F in N83C. The contour levels are the same as in Figure 1(d).
Right: map of Band~6 continuum emission (contour) superposed on H$\alpha$ flux map (color) in N83C.
The contour levels are 3, 4, 5, 7, 9, and 11 $\sigma$, where 1 $\sigma$ = 0.2 mJy beam$^{-1}$.
The synthesized beam is shown in the lower left corner.
}
\end{figure*}

\acknowledgments

We thank the anonymous referee for helpful comments, which significantly improved the manuscript.
This paper makes use of the following ALMA data: [ADS/JAO.ALMA\#2013.1.00212.S].
ALMA is a partnership of ESO (representing its member states), NSF (USA) and NINS (Japan), together with NRC (Canada),
NSC and ASIAA (Taiwan), and KASI (Republic of Korea), in cooperation with the Republic of Chile.
The Joint ALMA Observatory is operated by ESO, AUI/NRAO and NAOJ.
This work is based on observations made with the $Spitzer$ $Space$ $Telescope$,
which is operated by the Jet Propulsion Laboratory, California Institute of Technology under a contract with NASA.
This research made use of APLpy, an open-source plotting package for Python \citet{robitaille2012}.
This work was supported by NAOJ ALMA Scientific Research Grant Number 2016-03B and JSPS KAKENHI (Grant Nos.\ 17K14251, 22244014, 23403001, and 26247026).
This work was also supported by the Mitsubishi Foundation. M.\ Meixner was supported by NSF grant 1312902.

%

\vspace{5mm}

\end{document}